# Nonlinear Programming of Low-Thrust Multi-Rendezvous Trajectories Using Analytical Hessian

An-Yi Huang*

*State Key Laboratory of Astronautic Dynamics, Xi'an 710000, China*

Ya-Zhong Luo†

*National University of Defense Technology, Changsha, 410073, China*

## I. Introduction

The asteroids in the main belt hold significant potential for exploration value due to their diverse resources [1]. Given the large number of asteroids and their small mass, economic considerations necessitate that spacecraft are typically designed to sequentially explore multiple targets after reaching the main asteroid belt, thereby reducing the marginal costs of launch and return and maximizing mission benefits. Therefore, the optimization of multi-target rendezvous trajectories has always been a key issue in the design of such missions.

For single-spacecraft rendezvous sequences with a limited number of targets, mixed-integer programming algorithms can be employed to directly construct a full optimization model incorporating rendezvous order, timing, and impulsive or low-thrust maneuvers [2,3]. However, as the number of targets increases, the problem complexity grows rapidly, rendering such methods computationally intractable within acceptable timeframes. Particularly for global optimization problems involving multiple spacecraft and large-scale candidate targets, most studies opt for hierarchical decomposition with reasonable simplifications [4–8]. These approaches first perform a global search on target selection and sequencing, where objective functions and constraints are evaluated using computationally efficient approximations. Subsequently, a refined re-optimization model is established for each single-spacecraft rendezvous sequence. Finally, the transfer trajectories and control laws between targets are solved in segments.

In the process of the above global optimization algorithms, it is necessary to repeatedly evaluate the fuel consumption, time, and other custom task metrics of a large number of different rendezvous sequences that are randomly generated or obtained through crossover and mutation of feasible solutions. Therefore, a high-precision and fast sequence cost evaluation algorithm is crucial for improving efficiency. When conducting a preliminary feasibility screening of rendezvous sequences, methods such as branch and bound, beam search, or some heuristic approaches are typically used to approximate the task cost [5,8,9] or to establish a large-scale database of feasible sequences [10], focusing on quickly narrowing down the search space. These methods are fast in computation, but the evaluation results often have some deviation from the actual metrics, necessitating the application of nonlinear programming (NLP) algorithms [4,11–14], mixed-integer evolutionary algorithms [3,8], etc., for re-planning the sequences. However, evolutionary algorithms are usually less efficient for feeding the re-planning results back into the global optimization process for sequence improvement. Recently, analytical gradient-guided NLP algorithms [12–14] have been proposed for impulsive rendezvous sequences, enabling rapid local refinement of feasible solutions with efficiency sufficient for integration into global optimization frameworks, achieving state-of-the-art (SOTA) results in multiple problems. However, the methods above are limited to impulsive rendezvous scenarios.

* Associate Professor, Trajectory Optimization Team; hay04@foxmail.com.

† Professor, College of Aerospace Science and Engineering, and Hunan Key Laboratory of Intelligent Planning and Simulation for Aerospace Missions, luoyz@nudt.edu.cn.

This Note addresses the low-thrust multi-asteroid rendezvous trajectory optimization problem, where the transfer time between consecutive targets is much shorter than the orbital period. In this case, the double-impulse Lambert's solution [15] can be regarded as the optimal multi-impulse solution [14]. For scenarios using low thrust, a rough approximation is to scale Lambert's solution by a thrust-loss factor as an equivalent velocity increment [16–18], which can be used for preliminary feasibility screening of large-scale rendezvous sequences. The approach proposed in [19] can quickly assess the feasibility and fuel consumption of a given low-thrust transfer via parameter regression based on sampled data. Neural network-based approximators [20–22] have also been proposed, which are efficient for $\Delta v$ evaluation but time-consuming for data sampling and network training. Furthermore, the aforementioned methods cannot directly obtain gradients of $\Delta v$ with respect to rendezvous timing, transfer duration, and initial mass. As analytical gradients—particularly second-order gradients— are known to significantly improve the efficiency of NLP algorithms, their derivation constitutes the primary focus of this study.

The major contribution of this Note lies in proposing the analytical expression of first-order and second-order gradients of the Lambert's method-based low-thrust $\Delta v$ approximation, which are applied to derive the analytical Hessian matrix for nonlinear programming of multi-rendezvous trajectories. Simulation results indicate the analytical gradients are close to those obtained using the central differential method. Building upon the analytical first-order gradient and Hessian matrix, the sequential quadratic programming (SQP) algorithm exhibits local quadratic convergence for both fuel-optimal and time-optimal problems, significantly reducing computational costs in terms of objective function evaluations compared to approximate Hessian methods such as BFGS. The results submitted by the top three teams [4,18,23] in the 12th Global Trajectory Optimization Competition (GTOC12) [17] are then used as initial values to test the proposed algorithm with a more complicated objective function, which also achieves improvements in metrics with high computational efficiency. The high efficiency of the algorithm (e.g., solving a 9-target sequence in under 0.02 seconds) makes it practical for integration into existing multi-spacecraft multi-target global trajectory optimization frameworks.

The remainder of this Note is organized as follows. Section II formulates the multi-asteroid rendezvous sequence optimization problem. Section III presents the Lambert's solution-based iterative $\Delta v$ approximator of low-thrust rendezvous and the analytical gradients. Section IV details the NLP algorithm for optimizing sequences under various objectives and constraints. Section V validates the accuracy and computational efficiency of the approximate equivalent $\Delta v$ and its gradients, followed by verification of the multi-asteroid sequence NLP algorithm's performance. Finally, Section VI draws the conclusion.

## II. Problem Description

This Note focuses on the optimization of feasible low-thrust multi-asteroid rendezvous sequences obtained from prior searches, aiming to reduce the mission time or improve the fuel efficiency. The sequence search algorithm itself is not within the scope of this study. The problem is similar to the impulsive scenes that have been studied in [12,14], and the difference is the use of low-thrust propulsion.

In a single-spacecraft multi-rendezvous sequence with fixed targets, the spacecraft departs from an initial target, visits each target orbit in a prescribed order, and is allowed to wait for a specified duration after each rendezvous. The orbital motion of the spacecraft is expressed by

$$\begin{cases} \dot{\mathbf{r}} = \mathbf{v} \\ \dot{\mathbf{v}} = -\dfrac{\mu}{r^3}\mathbf{r} + \dfrac{F}{m}\boldsymbol{\alpha} \\ \dot{m} = -\dfrac{F}{I_{sp} g_e} \end{cases}, \tag{1}$$

where $\mathbf{r}$ is the position vector, $\mathbf{v}$ is the velocity vector, m is the mass, $I_{sp}$ is the specific impulse, $\boldsymbol{\alpha}$ is the normalized thrust direction, and $F$ is the thrust magnitude.

When the target number is $n$, the trajectory consists of $n$-1 pairwise rendezvous segments. The arrival epoch at the first target is limited to be no earlier than $t_0$ and the arrival epoch at the last target is limited to be not later than $t_f$. In this Note, an $n$-dimensional normalized vector $0 < x_i \le 1 (i = 0,1...n-1)$ is used to express the transfer durations of each transfer (denoted by $\Delta t_i$, where $\Delta t_0$ is the wait time before departing the first target and $\Delta t_i, 1 \le i \le n-1$ is the transfer duration between the $i^{\text{th}}$ and $i+1^{\text{th}}$ targets) and the arrival epochs at each target (denoted by $t_i$). The relationship between $\Delta t_i$, $t_i$, and $x_i$ is expressed by

$$\begin{cases} \Delta t_i = (\Delta t_{\max} - \Delta t_{\min}) x_i + \Delta t_{\min} \\ t_i = t_0 + \sum_{j=0}^{i} (\Delta t_j + \Delta t_{stay}) \end{cases}, \tag{2}$$

where $\Delta t_{\max}$ and $\Delta t_{\min}$ are the maximum and minimum transfer durations, respectively. $\Delta t_{stay}$ is the wait time after each rendezvous.

For the time-optimal problem with fuel constraint, the objective function is

$$J = t_n \tag{3}$$

The constraint is that the residual mass after rendezvousing with all targets ( $m_n$ ) must exceed the spacecraft's dry mass ( $m_{dry}$ ), as expressed by

$$m_{dry} - m_n \le 0, \tag{4}$$

where $m_n$ can be calculated recursively by starting from the initial mass ( $m_0$ ) using the expression as

$$m_i = m_{i-1} e^{\frac{-\Delta v_i}{I_{sp}}} - \Delta m_{kit}, \tag{5}$$

where $m_i$ denotes the mass after the $i^{\text{th}}$ rendezvous. $\Delta m_{kit}$ is the mass of the toolkit that may be released from the spacecraft to the asteroid (it can also be negative, which represents the mass collected from the target asteroid and added to the spacecraft). $\Delta v_j$ is the equivalent velocity increment of the $i^{\text{th}}$ transfer.

For the fuel-optimal problem with a limited mission completion time, the objective function is the fuel consumption as

$$J = m_0 - m_n \tag{6}$$

The constraint is expressed by

$$t_n - t_f \le 0 \tag{7}$$

Both the fuel-optimal and time-optimal problems must ensure each transfer is feasible. The constraint can be expressed by

$$\Delta v_i - \int_0^{\Delta t_i} \frac{F}{m} \mathrm{d}t \le 0 , \tag{8}$$

where $\int_0^{\Delta t_i} \frac{F}{m} \mathrm{d}t$ denotes the maximum velocity accumulated in $\Delta t_i$ when the thrust keeps on.

The sequential quadratic programming (SQP) algorithm [24] is adopted to solve this class of problems. Since evaluating the objective function requires repeated $\Delta v_i$ computations with different orbits and transfer durations, a balance between accuracy and efficiency is critical. SQP is proved to achieve higher efficiency with analytical gradients. If analytical second-order derivatives—specifically the Hessian matrix—can be obtained, it will further enable the iterative process to achieve quadratic convergence. Thus, this Note aims to develop a low-thrust $\Delta v$ approximator for which the first-order and second-order gradients with respect to the decision variables can be analytically calculated, serving as the underlying function for sequence optimization. The details are presented in the following sections.

## III. $\Delta v$ Approximator of Low-Thrust Rendezvous and Analytical Gradients

### A. Lambert-Based Equivalent $\Delta v$ Approximation

Although existing Lambert-based methods have been proposed to estimate velocity increments of low-thrust transfers, this section reestablishes a correspondence between the Lambert's double-impulse solution and the low-thrust trajectory to facilitate the derivation of analytical gradients. Given the orbital elements of the departure and arrival asteroids as well as the start epoch $t_0$ and flight duration $\Delta t$, the Lambert's method [15] can be employed to calculate the two impulses at the start and end epochs for orbit rendezvous. In this Note, the impulsive velocity increment is denoted by a function $f$ :

$$f(t_0, \Delta t) = f_1(t_0, \Delta t) + f_2(t_0, \Delta t) , \tag{9}$$

where $f_1$ and $f_2$ are the magnitudes of the impulses. Ref. [25] has provided the gradients of $f_1$ and $f_2$ with respect to $t_0$ and $\Delta t$ (denoted by $\partial f_1 / \partial t_0$ , $\partial f_1 / \partial \Delta t$ , $\partial f_2 / \partial t_0$ , and $\partial f_2 / \partial \Delta t$ ) with negligible additional computational cost.

For orbit rendezvous using continuous low-thrust propulsion, this Note assumes that the optimal control law follows a three-phase structure (thrust-on/off/on) similar to the Lambert's solution, where each thrusting phase can be approximated as an impulse applied at its midpoint. Let $\Delta t_1$ and $\Delta t_2$ denote the durations of the first and last thrusting arcs, respectively. When $t_0$ and $\Delta t$ are given, the low-thrust equivalent velocity increment can be approximated by the Lambert's double-impulse solution with a departure time of $t_0+\Delta t_1/2$ and a transfer duration of $\Delta t-(\Delta t_1+\Delta t_2)/2$, (i.e., $f(t_0+\Delta t_1/2, \Delta t-(\Delta t_1+\Delta t_2)/2)$ ). Here, $\Delta t_1$ and $\Delta t_2$ must satisfy the expression as

$$\begin{aligned} \Delta t_1 &= f_1(t_0+\frac{\Delta t_1}{2}, \Delta t-\frac{\Delta t_1}{2}-\frac{\Delta t_2}{2})/a \\ \Delta t_2 &= f_2(t_0+\frac{\Delta t_1}{2}, \Delta t-\frac{\Delta t_1}{2}-\frac{\Delta t_2}{2})/a \end{aligned}, \tag{10}$$

where $a$ is the average acceleration during the transfer, approximately calculated as

$$a=\frac{F}{(m_0+m_f)/2}=\frac{F}{m_0}\frac{2}{(1+e^{\frac{-f}{I_{sp}g_e}})}, \tag{11}$$

where $m_0$ is the mass before the transfer, and is the gravitational acceleration at the surface of the Earth's equator. In subsequent calculations, we denote $k_a=\frac{2}{(1+e^{\frac{-f}{I_{sp}g_e}})}$ as a constant.

Let $g_1=f_1(t_0+\frac{\Delta t_1}{2}, \Delta t-\frac{\Delta t_1}{2}-\frac{\Delta t_2}{2})$ and $g_2=f_2(t_0+\frac{\Delta t_1}{2}, \Delta t-\frac{\Delta t_1}{2}-\frac{\Delta t_2}{2})$ represent the equivalent velocity increments for the two thrusting arcs. Eq. (10) can alternatively be interpreted as follows: Since the beginning time is fixed to $t_0$, the first thrusting arc requires a certain duration to accumulate the orbital modification effect equivalent to an impulse, thereby delaying the epoch of the first equivalent impulsive by $\Delta t_1/2$. Similarly, the terminal time of the second thrusting arc is fixed to $t_0+\Delta t$, advancing its equivalent impulse epoch by $\Delta t_2/2$. The total equivalent velocity increment can be expressed by a function $g(t_0, \Delta t)$:

$$g(t_0, \Delta t)=g_1(t_0, \Delta t)+g_2(t_0, \Delta t) \tag{12}$$

Typically, when the transfer duration is much less than the orbital period, the double-impulse velocity increment of Lambert's solution increases monotonically with decreasing transfer time $\Delta t$ [14]. Thus, $\Delta t_1$ and $\Delta t_2$ can be computed via an iterative procedure:

First, calculate $f_1$ and $f_2$ via Eq. (9) by inputting $\Delta t$ and $t_0$, and then calculate $k_a$ by Eq. (11). Second, if $\frac{f_1+f_2}{a}>\Delta t$ holds, the low-thrust rendezvous is infeasible, and the iteration terminates. Third, modify the equivalent departure time to $t_0+\frac{f_1}{2a}$ and the equivalent transfer duration to $\Delta t-\frac{f_1+f_2}{2a}$, then recompute $f_1(t_0+\frac{\Delta t_1}{2},\Delta t-\frac{\Delta t_1}{2}-\frac{\Delta t_2}{2})$ and $f_2(t_0+\frac{\Delta t_1}{2},\Delta t-\frac{\Delta t_1}{2}-\frac{\Delta t_2}{2})$ via Eq. (9). Fourth, if the change in $f_1+f_2$ before and after the input modification falls below a given threshold $\varepsilon$, exit the iteration and return $f_1+f_2$ as approximate $g(t_0,\Delta t)$; Otherwise, update $k_a$ and return to the second step for the next iteration.

**B. First-Order Gradients of $\Delta v$**

To obtain analytical expressions of the gradients, this Note ignores the gradients of $k_a$ and considers it as a constant. A brief justification is provided below. Take $\frac{\partial k_a}{\partial t_0}$ as an example, its expression is

$$\frac{\partial k_a}{\partial t_0}=\frac{2}{I_{sp}g(1+e^{\frac{-g_1-g_2}{I_{sp}g_e}})^2}e^{\frac{-g_1-g_2}{I_{sp}g_e}}(\frac{\partial g_1}{\partial t_0}+\frac{\partial g_2}{\partial t_0}), \tag{13}$$

where $e^{\frac{-g_1-g_2}{I_{sp}g_e}}$ is close to 1 when the fuel consumption of a single transfer is small (i.e., when $\frac{g}{I_{sp}g_e}=0.1$, $e^{\frac{-g_1-g_2}{I_{sp}g_e}}$ is 1.1). $\frac{\partial k_a}{\partial t_0}/(\frac{\partial g_1}{\partial t_0}+\frac{\partial g_2}{\partial t_0})$ is close to $\frac{1}{2I_{sp}g_e}$, which is less than 1.3e-5 when $I_{sp}$ is 4,000 s. Thus, assuming $\frac{\partial k_a}{\partial t_0}=0$ is reasonable.

Then, according to Eq. (12), the gradients of $g_1$ and $g_2$ to $t_0$ are

$$\begin{aligned}\frac{\partial g_1}{\partial t_0}&=\frac{\partial f_1}{\partial(t_0+\frac{g_1}{2a})}(1+\frac{\partial g_1}{2a\partial t_0})+\frac{\partial f_1}{\partial(\Delta t-\frac{g_1+g_2}{2a})}(-\frac{\partial g_1}{2a\partial t_0}-\frac{\partial g_2}{2a\partial t_0})\\ \frac{\partial g_2}{\partial t_0}&=\frac{\partial f_2}{\partial(t_0+\frac{g_1}{2a})}(1+\frac{\partial g_1}{2a\partial t_0})+\frac{\partial f_2}{\partial(\Delta t-\frac{g_1+g_2}{2a})}(-\frac{\partial g_1}{2a\partial t_0}-\frac{\partial g_2}{2a\partial t_0})\end{aligned}, \tag{14}$$

where $\frac{\partial f_1}{\partial(t_0+\frac{g_1}{2a})}$, $\frac{\partial f_1}{\partial(\Delta t-\frac{g_1+g_2}{2a})}$, $\frac{\partial f_2}{\partial(t_0+\frac{g_1}{2a})}$, and $\frac{\partial f_2}{\partial(\Delta t-\frac{g_1+g_2}{2a})}$ denote the gradient of the Lambert's solution with respect to the beginning epoch and transfer duration that is calculated by the method in [25] (using the low-thrust equivalent inputs). Eq. (14) can be rewritten in matrix form as

$$\begin{bmatrix}1-\frac{\partial f_1}{2a\partial(t_0+\frac{g_1}{2a})}+\frac{\partial f_1}{2a\partial(\Delta t-\frac{g_1+g_2}{2a})} & \frac{\partial f_1}{2a\partial(\Delta t-\frac{g_1+g_2}{2a})}\\ -\frac{\partial f_2}{2a\partial(t_0+\frac{g_1}{2a})}+\frac{\partial f_2}{2a\partial(\Delta t-\frac{g_1+g_2}{2a})} & 1+\frac{\partial f_2}{2a\partial(\Delta t-\frac{g_1+g_2}{2a})}\end{bmatrix}\begin{bmatrix}\frac{\partial g_1}{\partial t_0}\\ \frac{\partial g_2}{\partial t_0}\end{bmatrix}=\begin{bmatrix}\frac{\partial f_1}{\partial(t_0+\frac{g_1}{2a})}\\ \frac{\partial f_2}{\partial(t_0+\frac{g_1}{2a})}\end{bmatrix} \tag{15}$$

In the same manner, the gradients of $g_1$ and $g_2$ to $\Delta t$ are expressed by

$$\begin{bmatrix} 1-\dfrac{\partial f_1}{2a\partial(t_0+\frac{g_1}{2a})}+\dfrac{\partial f_1}{2a\partial(\Delta t-\frac{g_1+g_2}{2a})} & \dfrac{\partial f_1}{2a\partial(\Delta t-\frac{g_1+g_2}{2a})} \\ -\dfrac{\partial f_2}{2a\partial(t_0+\frac{g_1}{2a})}+\dfrac{\partial f_2}{2a\partial(\Delta t-\frac{g_1+g_2}{2a})} & 1+\dfrac{\partial f_2}{2a\partial(\Delta t-\frac{g_1+g_2}{2a})} \end{bmatrix} \begin{bmatrix} \dfrac{\partial g_1}{\partial \Delta t} \\ \dfrac{\partial g_2}{\partial \Delta t} \end{bmatrix} = \begin{bmatrix} \dfrac{\partial f_1}{\partial(\Delta t-\frac{g_1+g_2}{2a})} \\ \dfrac{\partial f_2}{\partial(\Delta t-\frac{g_1+g_2}{2a})} \end{bmatrix} \tag{16}$$

Additionally, since the equivalent velocity increments depend on thrust acceleration, the gradients of $g_1$ and $g_2$ to the initial mass $m_0$ are required for sequence planning. The expression is

$$\begin{bmatrix} 1-\dfrac{\partial f_1}{2a\partial(t_0+\frac{g_1}{2a})}+\dfrac{\partial f_1}{2a\partial(\Delta t-\frac{g_1+g_2}{2a})} & \dfrac{\partial f_1}{2a\partial(\Delta t-\frac{g_1+g_2}{2a})} \\ -\dfrac{\partial f_2}{2a\partial(t_0+\frac{g_1}{2a})}+\dfrac{\partial f_2}{2a\partial(\Delta t-\frac{g_1+g_2}{2a})} & 1+\dfrac{\partial f_2}{2a\partial(\Delta t-\frac{g_1+g_2}{2a})} \end{bmatrix} \begin{bmatrix} \dfrac{\partial g_1}{\partial m_0} \\ \dfrac{\partial g_2}{\partial m_0} \end{bmatrix} = \frac{1}{2k_aF} \begin{bmatrix} \dfrac{\partial f_1}{\partial(t_0+\frac{g_1}{2a})}g_1-\dfrac{\partial f_1}{\partial(\Delta t-\frac{g_1+g_2}{2a})}(g_1+g_2) \\ \dfrac{\partial f_2}{\partial(t_0+\frac{g_1}{2a})}g_1-\dfrac{\partial f_2}{\partial(\Delta t-\frac{g_1+g_2}{2a})}(g_1+g_2) \end{bmatrix} \tag{17}$$

Note that the left-hand matrices in Eqs. (15), (16), and (17) are identical and let **A** denote it. The gradients of $g_1$ and $g_2$ can be obtained by

$$\begin{aligned} \begin{bmatrix} \dfrac{\partial g_1}{\partial t_0} \\ \dfrac{\partial g_2}{\partial t_0} \end{bmatrix} &= A^{-1} \begin{bmatrix} \dfrac{\partial f_1}{\partial(t_0+\frac{g_1}{2a})} \\ \dfrac{\partial f_2}{\partial(t_0+\frac{g_1}{2a})} \end{bmatrix} \\ \begin{bmatrix} \dfrac{\partial g_1}{\partial \Delta t} \\ \dfrac{\partial g_2}{\partial \Delta t} \end{bmatrix} &= A^{-1} \begin{bmatrix} \dfrac{\partial f_1}{\partial(\Delta t-\frac{g_1+g_2}{2a})} \\ \dfrac{\partial f_2}{\partial(\Delta t-\frac{g_1+g_2}{2a})} \end{bmatrix} \\ \begin{bmatrix} \dfrac{\partial g_1}{\partial m_0} \\ \dfrac{\partial g_2}{\partial m_0} \end{bmatrix} &= \frac{A^{-1}}{2k_aF} \begin{bmatrix} \dfrac{\partial f_1}{\partial(t_0+\frac{g_1}{2a})}g_1-\dfrac{\partial f_1}{\partial(\Delta t-\frac{g_1+g_2}{2a})}(g_1+g_2) \\ \dfrac{\partial f_2}{\partial(t_0+\frac{g_1}{2a})}g_1-\dfrac{\partial f_2}{\partial(\Delta t-\frac{g_1+g_2}{2a})}(g_1+g_2) \end{bmatrix} \end{aligned} \tag{18}$$

**C. Second-Order Gradients of $\Delta v$**

Let $\begin{bmatrix} f_{1xx} & f_{1xy} \\ f_{1yx} & f_{1yy} \end{bmatrix} = \begin{bmatrix} \dfrac{\partial^2 f_1}{\partial^2 t_0} & \dfrac{\partial^2 f_1}{\partial t_0 \partial \Delta t} \\ \dfrac{\partial^2 f_1}{\partial \Delta t \partial t_0} & \dfrac{\partial^2 f_1}{\partial^2 \Delta t} \end{bmatrix}$ and $\begin{bmatrix} f_{2xx} & f_{2xy} \\ f_{2yx} & f_{2yy} \end{bmatrix} = \begin{bmatrix} \dfrac{\partial^2 f_2}{\partial^2 t_0} & \dfrac{\partial^2 f_2}{\partial t_0 \partial \Delta t} \\ \dfrac{\partial^2 f_2}{\partial \Delta t \partial t_0} & \dfrac{\partial^2 f_2}{\partial^2 \Delta t} \end{bmatrix}$ denote the second-order gradients of the Lambert's solution, which can be obtained by the method proposed in [] or the central-differential method. Taking the partial derivative of Eq. (15) with respect to $t_0$, and let $\frac{\partial g_1}{\partial t_0}, \frac{\partial g_1}{\partial \Delta t}, \frac{\partial g_1}{\partial m_0}, \frac{\partial g_2}{\partial t_0}, \frac{\partial g_2}{\partial \Delta t}, \frac{\partial g_2}{\partial m_0}$ be denoted as $g_{1x}, g_{1y}, g_{1m}, g_{2x}, g_{2y}, g_{2m}$ respectively for ease of formula notation, $\frac{\partial^2 g_1}{\partial \Delta t \partial \Delta t}$ and $\frac{\partial^2 g_2}{\partial \Delta t \partial \Delta t}$ are expressed by

$$A\begin{bmatrix} \frac{\partial^2 g_1}{\partial^2 t_0} \\ \frac{\partial^2 g_2}{\partial^2 t_0} \end{bmatrix} = \begin{bmatrix} f_{1xy}(-\frac{g_{1x}+g_{2x}}{2a}) + f_{1xx}(1+\frac{g_{1x}}{2a}) \\ f_{2xy}(-\frac{g_{1x}+g_{2x}}{2a}) + f_{2xx}(1+\frac{g_{1x}}{2a}) \end{bmatrix} - \frac{1}{2a}\begin{bmatrix} -f_{1xx}(1+\frac{g_{1x}}{2a}) - f_{1xy}(-\frac{g_{1x}+g_{2x}}{2a}) + f_{1yy}(-\frac{g_{1x}+g_{2x}}{2a}) + f_{1yx}(1+\frac{g_{1x}}{2a}) & f_{1yy}(-\frac{g_{1x}+g_{2x}}{2a}) + f_{1yx}(1+\frac{g_{1x}}{2a}) \\ -f_{2xx}(1+\frac{g_{1x}}{2a}) - f_{2xy}(-\frac{g_{1x}+g_{2x}}{2a}) + f_{2yy}(-\frac{g_{1x}+g_{2x}}{2a}) + f_{2yx}(1+\frac{g_{1x}}{2a}) & f_{2yy}(-\frac{g_{1x}+g_{2x}}{2a}) + f_{2yx}(1+\frac{g_{1x}}{2a}) \end{bmatrix}\begin{bmatrix} g_{1x} \\ g_{2x} \end{bmatrix} \tag{19}$$

In the same manner, $\frac{\partial^2 g_1}{\partial \Delta t \partial \Delta t}$ and $\frac{\partial^2 g_2}{\partial \Delta t \partial \Delta t}$, that can be obtained by taking the partial derivative of Eq. (15) with respect to $\Delta t$, are expressed by

$$A\begin{bmatrix} \frac{\partial^2 g_1}{\partial \Delta t \partial \Delta t} \\ \frac{\partial^2 g_2}{\partial \Delta t \partial \Delta t} \end{bmatrix} = \begin{bmatrix} f_{1yy}(1-\frac{g_{1y}+g_{2y}}{2a}) + f_{1yx}\frac{g_{1y}}{2a} \\ f_{2yy}(1-\frac{g_{1y}+g_{2y}}{2a}) + f_{2yx}\frac{g_{1y}}{2a} \end{bmatrix} - \frac{1}{2a}\begin{bmatrix} -f_{1xx}\frac{g_{1y}}{2a} - f_{1xy}(1-\frac{g_{1y}+g_{2y}}{2a}) + f_{1yy}(1-\frac{g_{1y}+g_{2y}}{2a}) + f_{1yx}\frac{g_{1y}}{2a} & f_{1yy}(1-\frac{g_{1y}+g_{2y}}{2a}) + f_{1yx}\frac{g_{1y}}{2a} \\ -f_{2xx}\frac{g_{1y}}{2a} - f_{2xy}(1-\frac{g_{1y}+g_{2y}}{2a}) + f_{2yy}(1-\frac{g_{1y}+g_{2y}}{2a}) + f_{2yx}\frac{g_{1y}}{2a} & f_{2yy}(1-\frac{g_{1y}+g_{2y}}{2a}) + f_{2yx}\frac{g_{1y}}{2a} \end{bmatrix}\begin{bmatrix} g_{1y} \\ g_{2y} \end{bmatrix} \tag{20}$$

$\frac{\partial^2 g_1}{\partial \Delta t \partial t_0} = \frac{\partial^2 g_1}{\partial t_0 \partial \Delta t}$ and $\frac{\partial^2 g_2}{\partial \Delta t \partial t_0} = \frac{\partial^2 g_2}{\partial t_0 \partial \Delta t}$, that can be obtained by taking the partial derivative of Eq. (15) with respect to $\Delta t$, are expressed by

$$A\begin{bmatrix} \frac{\partial^2 g_1}{\partial t_0 \partial \Delta t} \\ \frac{\partial^2 g_2}{\partial t_0 \partial \Delta t} \end{bmatrix} = \begin{bmatrix} f_{1xx}\frac{g_{1y}}{2a} + f_{1xy}(1-\frac{g_{1y}+g_{2y}}{2a}) \\ f_{2xx}\frac{g_{1y}}{2a} + f_{2xy}(1-\frac{g_{1y}+g_{2y}}{2a}) \end{bmatrix} - \frac{1}{2a}\begin{bmatrix} -f_{1xx}\frac{g_{1y}}{2a} - f_{1xy}(1-\frac{g_{1y}+g_{2y}}{2a}) + f_{1yx}\frac{g_{1y}}{2a} + f_{1yy}(1-\frac{g_{1y}+g_{2y}}{2a}) & f_{1yx}\frac{g_{1y}}{2a} + f_{1yy}(1-\frac{g_{1y}+g_{2y}}{2a}) \\ -f_{2xx}\frac{g_{1y}}{2a} - f_{2xy}(1-\frac{g_{1y}+g_{2y}}{2a}) + f_{2yx}\frac{g_{1y}}{2a} + f_{2yy}(1-\frac{g_{1y}+g_{2y}}{2a}) & f_{2yx}\frac{g_{1y}}{2a} + f_{2yy}(1-\frac{g_{1y}+g_{2y}}{2a}) \end{bmatrix}\begin{bmatrix} g_{1x} \\ g_{2x} \end{bmatrix} \tag{20}$$

Also, the second-order gradients related to the mass, $\frac{\partial^2 g_1}{\partial t_0 \partial m_0}$, $\frac{\partial^2 g_1}{\partial \Delta t \partial m_0}$, $\frac{\partial^2 g_1}{\partial^2 m_0}$, $\frac{\partial^2 g_2}{\partial t_0 \partial m_0}$, $\frac{\partial^2 g_2}{\partial \Delta t \partial m_0}$, and $\frac{\partial^2 g_2}{\partial^2 m_0}$, can be expressed by:

$$A\begin{bmatrix} \frac{\partial^2 g_1}{\partial^2 m_0} \\ \frac{\partial^2 g_2}{\partial^2 m_0} \end{bmatrix} = \frac{1}{2am_0}\begin{bmatrix} f_{1x}g_{1m} + f_{1xx}g_1\frac{\partial(\frac{g_1}{2a})}{\partial m_0} - f_{1xy}g_1(\frac{\partial(\frac{g_1+g_2}{2a})}{\partial m_0}) - f_{1y}(g_{1m}+g_{2m}) - f_{1yx}\frac{\partial(\frac{g_1}{2a})}{\partial m_0}(g_1+g_2) + f_{1yy}(\frac{\partial(\frac{g_1+g_2}{2a})}{\partial m_0})(g_1+g_2) \\ f_{2x}g_{1m} + f_{2xx}g_1\frac{\partial(\frac{g_1}{2a})}{\partial m_0} - f_{2xy}g_1(\frac{\partial(\frac{g_1+g_2}{2a})}{\partial m_0}) - f_{2y}(g_{1m}+g_{2m}) - f_{2yx}\frac{\partial(\frac{g_1}{2a})}{\partial m_0}(g_1+g_2) + f_{2yy}(\frac{\partial(\frac{g_1+g_2}{2a})}{\partial m_0})(g_1+g_2) \end{bmatrix} - \frac{\partial A}{\partial m}\begin{bmatrix} g_{1m} \\ g_{2m} \end{bmatrix}$$

$$A\begin{bmatrix} \frac{\partial^2 g_1}{\partial t_0 \partial m_0} \\ \frac{\partial^2 g_2}{\partial t_0 \partial m_0} \end{bmatrix} = \begin{bmatrix} f_{1xx}\frac{\partial(\frac{g_{1x}}{2a})}{\partial m_0} - f_{1xy}\frac{\partial(\frac{g_1+g_2}{2a})}{\partial m_0} \\ f_{2xx}\frac{\partial(\frac{g_{1x}}{2a})}{\partial m_0} - f_{2xy}\frac{\partial(\frac{g_1+g_2}{2a})}{\partial m_0} \end{bmatrix} - \frac{\partial A}{\partial m_0}\begin{bmatrix} g_{1x} \\ g_{2x} \end{bmatrix} \tag{21}$$

$$A\begin{bmatrix} \frac{\partial^2 g_1}{\partial \Delta t \partial m_0} \\ \frac{\partial^2 g_2}{\partial \Delta t \partial m_0} \end{bmatrix} = \begin{bmatrix} f_{1yx}\frac{\partial(\frac{g_{1x}}{2a})}{\partial m_0} - f_{1yy}\frac{\partial(\frac{g_1+g_2}{2a})}{\partial m_0} \\ f_{2yx}\frac{\partial(\frac{g_{1x}}{2a})}{\partial m_0} - f_{2yy}\frac{\partial(\frac{g_1+g_2}{2a})}{\partial m_0} \end{bmatrix} - \frac{\partial A}{\partial m_0}\begin{bmatrix} g_{1y} \\ g_{2y} \end{bmatrix}$$

where $\frac{\partial(\frac{g_1}{2a})}{\partial m_0} = \frac{1}{2a}\frac{\partial g_1}{\partial m_0} + \frac{g_1}{2am_0}$, $\frac{\partial(\frac{g_2}{2a})}{\partial m_0} = \frac{1}{2a}\frac{\partial g_2}{\partial m_0} + \frac{g_2}{2am_0}$, and $\frac{\partial A}{\partial m_0}$ is expressed by:

$$\frac{\partial A}{\partial m_0}=\frac{1}{2a}\begin{bmatrix} -f_{1xx}\frac{\partial(\frac{g_1}{2a})}{\partial m_0}+f_{1xy}\frac{\partial(\frac{g_1+g_2}{2a})}{\partial m_0}+f_{1yx}\frac{\partial(\frac{g_1}{2a})}{\partial m_0}-f_{1yy}\frac{\partial(\frac{g_1+g_2}{2a})}{\partial m_0} & f_{1yx}\frac{\partial(\frac{g_1}{2a})}{\partial m_0}-f_{1yy}\frac{\partial(\frac{g_1+g_2}{2a})}{\partial m_0} \\ -f_{2xx}\frac{\partial(\frac{g_1}{2a})}{\partial m_0}+f_{2xy}\frac{\partial(\frac{g_1+g_2}{2a})}{\partial m_0}+f_{2yx}\frac{\partial(\frac{g_1}{2a})}{\partial m_0}-f_{2yy}\frac{\partial(\frac{g_1+g_2}{2a})}{\partial m_0} & f_{2yx}\frac{\partial(\frac{g_1}{2a})}{\partial m_0}-f_{2yy}\frac{\partial(\frac{g_1+g_2}{2a})}{\partial m_0} \end{bmatrix} + \frac{1}{2am_0}\begin{bmatrix} -f_{1x}+f_{1y} & f_{1y} \\ -f_{2x}+f_{2y} & f_{2y} \end{bmatrix} \tag{22}$$

With the proposed analytical gradient formulations, a nonlinear programming algorithm (detailed in the next section) using analytical gradients and Hessian matrices can be established to optimize the multi-asteroid rendezvous sequences.

## IV. Nonlinear Programming Algorithm for Low-thrust Multi-Asteroid Sequences

As described in Section II, the decision variables are $x_i, i=0,1...n-1$. The objective and constraint functions are defined as follows: Eqs. (3) and (4) for time-optimal problems, and Eqs (5) and (6) for fuel-optimal problems. Additionally, the feasibility assessment (Eq. (8)) for both problems can be rewritten to

$$\Delta v_i - \frac{k_{a,i}T}{m_i}\Delta t_i \le 0, \tag{23}$$

where the subscripts '$i$' in $\Delta v_i$, $k_{a,i}$ and $m_i$ represent the variables correspond to the $i^{\text{th}}$ transfer in the sequence. Note that a feasibility assessment is required for every transfer; therefore, there are $n$-1 different constraint functions.

### A. First-Order Gradients of Objective and Constraint Functions

The first-order gradients of objective and constraint functions with respect to the decision variables are derived as follows. First, according to Eq. (2), the gradients of $t_i$ and $\Delta t_i$ to each component in $\mathbf{x}$ are

$$\frac{\partial t_i}{\partial x_j}=\begin{cases}\Delta t_{\max}-\Delta t_{\min}, i>j \\ 0, i\le j\end{cases}$$
$$\frac{\partial \Delta t_i}{\partial x_j}=\begin{cases}\Delta t_{\max}-\Delta t_{\min}, i=j \\ 0, i\ne j\end{cases} \tag{24}$$

Then, the gradients of $\Delta v_i$ to each component in $\mathbf{x}$ are

$$\frac{\partial \Delta v_i}{\partial x_j}=\frac{\partial \Delta v_i}{\partial t_i}\frac{\partial t_i}{\partial x_j}+\frac{\partial \Delta v_i}{\partial \Delta t_i}\frac{\partial \Delta t_i}{\partial x_j}+\frac{\partial \Delta v_i}{\partial m_i}\frac{\partial m_i}{\partial x_j}, \tag{25}$$

where $\frac{\partial \Delta v_i}{\partial t_i}$, $\frac{\partial \Delta v_i}{\partial \Delta t_i}$, and $\frac{\partial \Delta v_i}{\partial m_i}$ are obtained by Eq. (18). $\frac{\partial m_i}{\partial x_j}$ is expressed by

$$\frac{\partial m_i}{\partial x_j}=\frac{\partial m_{i-1}}{\partial x_j}e^{\frac{-\Delta v_{i-1}}{I_{sp}g_e}}+\frac{-m_{i-1}}{I_{sp}g_e}\frac{\partial \Delta v_{i-1}}{\partial x_j} \tag{26}$$

As seen in Eqs. (25) and (26), $\frac{\partial m_i}{\partial x_j}$ and $\frac{\partial \Delta v_i}{\partial x_j}$ are coupled. Given the initial spacecraft mass $m_0$ ($\frac{\partial m_0}{\partial x_j}$ =0), one can first calculate $\frac{\partial \Delta v_1}{\partial x_j}$, and then calculate $\frac{\partial m_1}{\partial x_j}$ by substituting $\frac{\partial \Delta v_1}{\partial x_j}$ into Eq. (26). $\frac{\partial m_i}{\partial x_j}$ and $\frac{\partial \Delta v_i}{\partial x_j}$ ($j$ > 1) can then be obtained consequently.

Thus, for time-optimal planning, the first-order gradient of the objective function Eq. (3) is $\sum_{i=0}^{n-1}\frac{\partial \Delta t_i}{\partial x_j}$ and the first-order gradient of the constraint function Eq. (4) is $\frac{-\partial m_n}{\partial x_j}$. For fuel-optimal planning, the first-order gradient of the objective function Eq. (5) is $\frac{-\partial m_n}{\partial x_j}$ and the first-order gradient of the constraint function Eq. (6) is $\sum_{i=0}^{n-1}\frac{\partial \Delta t_i}{\partial x_j}$.

Additionally, the gradient of the constraint function Eq. (23) is

$$\frac{\partial \Delta v_i}{\partial x_j}-a_i\frac{\partial \Delta t_i}{\partial x_j}-\frac{\partial a_i}{\partial x_j}\Delta t_i=\frac{\partial \Delta v_i}{\partial x_j}-\frac{k_aF}{m_{i-1}}\frac{\partial \Delta t_i}{\partial x_j}+\frac{k_aF}{m_{i-1}^{\ 2}}\frac{\partial m_{i-1}}{\partial x_j}\Delta t_i \tag{27}$$

**B. Second-Order Gradients of Objective and Constraint Functions**

The second-order gradients of $\Delta v_i$ are expressed by

$$\begin{aligned}
\frac{\partial \Delta v_i}{\partial x_k \partial x_j}&=\frac{\partial\left(\frac{\partial \Delta v_i}{\partial t_i}\frac{\partial t_i}{\partial x_j}+\frac{\partial \Delta v_i}{\partial \Delta t_i}\frac{\partial \Delta t_i}{\partial x_j}+\frac{\partial \Delta v_i}{\partial m_i}\frac{\partial m_i}{\partial x_j}\right)}{\partial x_k}\\
&=\frac{\partial\left(\frac{\partial \Delta v_i}{\partial t_i}\right)}{\partial x_k}\frac{\partial t_i}{\partial x_j}+\frac{\partial \Delta v_i}{\partial t_i}\frac{\partial^2 t_i}{\partial x_k \partial x_j}+\frac{\partial\left(\frac{\partial \Delta v_i}{\partial \Delta t_i}\right)}{\partial x_k}\frac{\partial \Delta t_i}{\partial x_j}+\frac{\partial \Delta v_i}{\partial \Delta t_i}\frac{\partial^2 \Delta t_i}{\partial x_k \partial x_j}+\frac{\partial\left(\frac{\partial \Delta v_i}{\partial m_i}\right)}{\partial x_k}\frac{\partial m_i}{\partial x_j}+\frac{\partial \Delta v_i}{\partial m_i}\frac{\partial^2 m_i}{\partial x_k \partial x_j}\\
&=\begin{bmatrix}\frac{\partial t_i}{\partial x_k} & \frac{\partial \Delta t_i}{\partial x_k} & \frac{\partial m_i}{\partial x_k}\end{bmatrix}\begin{bmatrix}\frac{\partial^2 \Delta v_i}{\partial^2 t_i} & \frac{\partial^2 \Delta v_i}{\partial t_i \partial \Delta t_i} & \frac{\partial^2 \Delta v_i}{\partial t_i \partial m_i}\\ \frac{\partial^2 \Delta v_i}{\partial \Delta t_i \partial t_i} & \frac{\partial^2 \Delta v_i}{\partial^2 \Delta t_i} & \frac{\partial^2 \Delta v_i}{\partial \Delta t_i \partial m_i}\\ \frac{\partial^2 \Delta v_i}{\partial m_i \partial t_i} & \frac{\partial^2 \Delta v_i}{\partial m_i \partial \Delta t_i} & \frac{\partial^2 \Delta v_i}{\partial^2 m_i}\end{bmatrix}\begin{bmatrix}\frac{\partial t_i}{\partial x_j}\\ \frac{\partial \Delta t_i}{\partial x_j}\\ \frac{\partial m_i}{\partial x_j}\end{bmatrix}+\frac{\partial \Delta v_i}{\partial m_i}\frac{\partial^2 m_i}{\partial x_k \partial x_j}
\end{aligned} \tag{28}$$

where we use the feature that $\frac{\partial(\frac{\partial t_i}{\partial x_j})}{\partial x_k} = \frac{\partial(\frac{\partial \Delta t_i}{\partial x_j})}{\partial x_k} = 0$, $\begin{bmatrix} \frac{\partial^2 \Delta v_i}{\partial^2 t_i} & \frac{\partial^2 \Delta v_i}{\partial t_i \partial \Delta t_i} & \frac{\partial^2 \Delta v_i}{\partial t_i \partial m_i} \\ \frac{\partial^2 \Delta v_i}{\partial \Delta t_i \partial t_i} & \frac{\partial^2 \Delta v_i}{\partial^2 \Delta t_i} & \frac{\partial^2 \Delta v_i}{\partial \Delta t_i \partial m_i} \\ \frac{\partial^2 \Delta v_i}{\partial m_i \partial t_i} & \frac{\partial^2 \Delta v_i}{\partial m_i \partial \Delta t_i} & \frac{\partial^2 \Delta v_i}{\partial^2 m_i} \end{bmatrix}$ is calculated via Eqs. (19) to (22), and $\frac{\partial^2 m_i}{\partial x_k \partial x_j}$ is the second-order gradients of mass when arriving at each target expressed by

$$\frac{\partial^2 m_i}{\partial x_k \partial x_j} = \frac{\partial^2 m_{i-1}}{\partial x_j \partial x_k} e^{\frac{-\Delta v_{i-1}}{I_{sp} g_e}} + \frac{\partial m_{i-1}}{\partial x_j} e^{\frac{-\Delta v_{i-1}}{I_{sp} g_e}} \frac{-1}{I_{sp} g_e} \frac{\partial \Delta v_{i-1}}{\partial x_k} - \frac{1}{I_{sp} g_e} \frac{\partial m_{i-1}}{\partial x_k} e^{\frac{-\Delta v_{i-1}}{I_{sp} g_e}} \frac{\partial \Delta v_{i-1}}{\partial x_j} + \frac{m_{i-1}}{(I_{sp} g_e)^2} e^{\frac{-\Delta v_{i-1}}{I_{sp} g_e}} \frac{\partial \Delta v_{i-1}}{\partial x_k} \frac{\partial \Delta v_{i-1}}{\partial x_j} - \frac{m_{i-1}}{I_{sp} g_e} e^{\frac{-\Delta v_{i-1}}{I_{sp} g_e}} \frac{\partial^2 \Delta v_{i-1}}{\partial x_k \partial x_j}$$

(29)

As $\frac{\partial^2 m_i}{\partial x_k \partial x_j}$ and $\frac{\partial \Delta v_i}{\partial x_k \partial x_j}$ are coupled and $\frac{\partial^2 m_0}{\partial x_k \partial x_j} = 0$ (the initial mass is fixed), we can first start from $i$=0 and calculate $\frac{\partial \Delta v_0}{\partial x_k \partial x_j}$. Then, $\frac{\partial^2 m_i}{\partial x_k \partial x_j}$ and $\frac{\partial \Delta v_i}{\partial x_k \partial x_j}$ can be alternately calculated from $i$=1 to $i = n$-1. Finally, $\frac{\partial^2 m_n}{\partial x_k \partial x_j}$ can be obtained.

Thus, for time-optimal planning, the second-order gradient of objective function Eq. (3) is $\sum_{j=0}^{n-1} \frac{\partial^2 \Delta t_j}{\partial x_j \partial x_k} = 0$ and the second-order gradient of the constraint function Eq. (4) is $-\frac{\partial^2 m_n}{\partial x_k \partial x_j}$. For fuel-optimal planning, the second-order gradient of objective function Eq. (5) is $-\frac{\partial^2 m_n}{\partial x_k \partial x_j}$ and the second-order gradient of constraint function Eq. (6) is $\sum_{j=0}^{n-1} \frac{\partial^2 \Delta t_j}{\partial x_j \partial x_k} = 0$. The second-order gradient of the constraint function Eq. (23) is

$$\frac{\partial^2 \Delta v_i}{\partial x_j \partial x_k} + \frac{k_a F}{m_{j-1}^2} \frac{\partial \Delta t_i}{\partial x_j} \frac{\partial m_{i-1}}{\partial x_k} + \frac{k_a F}{m_{j-1}^2} \frac{\partial m_{i-1}}{\partial x_j} \frac{\partial \Delta t_i}{\partial x_k} - \frac{2 k_a F}{m_{j-1}^3} \frac{\partial m_{i-1}}{\partial x_j} \frac{\partial m_{i-1}}{\partial x_k} \Delta t_i + \frac{k_a F}{m_{j-1}^2} \frac{\partial^2 m_{i-1}}{\partial x_j \partial x_k} \Delta t_i \tag{30}$$

### C. SQP Algorithm Using Analytical Gradients and Hessian Matrix

SQP iteratively solves nonlinear problems by generating quadratic programming (QP) subproblems that approximate the Lagrangian at the current solution (denoted by $\mathbf{x}_k$), computing a search direction $\mathbf{d}_k$ via QP solution, conducting a line search for step size $\alpha$ to minimize a predefined merit function, then updating $\mathbf{x}_{k+1}$ =

$\mathbf{x}_k + \alpha\mathbf{d}_k$ until convergence thresholds for variable increments and objective function changes are met. Conventional SQP implementations (e.g., MATLAB's 'fmincon' or SNOPT toolkit) typically employ approximate Hessian update strategies such as BFGS to achieve super-linear convergence, as analytical Hessian expressions are generally unavailable for most problems. However, this section derives the closed-form expression of the Hessian matrix for multi-rendezvous sequence optimization, eliminating the need for approximation schemes and thereby enabling quadratic convergence.

The QP subproblem for the multi-rendezvous sequence optimization problem can be expressed by

$$\begin{aligned} &\min_{\mathbf{d}} \ \mathbf{d}_k{}^T \nabla^2 L(x_k,\mu)\mathbf{d}_k + \nabla J(x_k)^T \mathbf{d}_k \\ &\text{s.t. } \nabla g_i(x_k)^T \mathbf{d}_k + g_i(x_k)^T \le 0, i = 0,2...n-1 \end{aligned} \tag{31}$$

where $J$ represents the objective function (Eq. (3) for time-optimal or Eq.(6) for fuel-optimal), $g_0$ represents the constraint function (Eq. (4) for time-optimal or Eq.(7) for fuel-optimal), represents the low-thrust feasibility constraints of the $n$-1 transfer legs, $L(x_k,\mu) = f(x_k) + \mu^{\mathrm{T}} g(x_k)$ represents the Lagrangian, and $\mu^{\mathrm{T}} = [\mu_0, \mu_1, ..., \mu_{n-1}]^T$ denotes the Lagrange multipliers. The expressions of $\nabla f(x_k)^T$ and $\nabla g_i(x_k)^T$ have been proposed in Section IV.B. $\nabla^2 L(x_k,\mu)$ is the Hessian matrix, and the expression for the time-optimal problem is

$$\nabla^2 L(x_k,\mu) = -\frac{\partial^2 m_n}{\partial x_k \partial x_j} + \sum_{i=1}^{n-1} \frac{\mu_i \partial^2 g_i}{\partial x_k \partial x_j} \tag{32}$$

For the fuel-optimal problem, the expression of $\nabla^2 L(x_k,\mu)$ is

$$\nabla^2 L(x_k,\mu) = -\frac{\mu_0 \partial^2 m_n}{\partial x_k \partial x_j} + \sum_{i=1}^{n-1} \frac{\mu_i \partial^2 g_i}{\partial x_k \partial x_j} \tag{33}$$

Eq. (31) can be easily solved via traditional QP solvers. When $\mathbf{d}_k$ is obtained, the merit function for line search to determine $\alpha \in (0,1]$ is

$$\Phi(\alpha) = f(x_k + \alpha\mathbf{d}_k) + \sum_{i=0}^{n-1} \kappa g_i(x_k + \alpha\mathbf{d}_k) \tag{34}$$

where $\kappa$ is a predefined coefficient. Since the analytical gradient of $\Phi(\alpha)$ can be calculated by the gradients of $f$ and $g$, the Wolfe condition line search[26] is employed to solve $\alpha$ , typically requiring about five objective function evaluations. Collectively, this section establishes an SQP algorithm for multi-rendezvous sequence optimization based on analytic gradients and Hessian.

## V. Simulation Results

This section verifies the effectiveness of the proposed method through multi-asteroid rendezvous missions in the main asteroid belt. First, we randomly generate rendezvous datasets with different initial and target orbital elements to validate the proposed $\Delta v$ and gradient calculations using the benchmark results obtained by the indirect method [27]. Then, three multi-rendezvous cases with different objective functions (time-optimal, fuel-optimal, and a weighted-sum time metric designed in GTOC12 [17]) are used to test the SQP's efficiency for local improvement of feasible sequences.

### A. Accuracy Verification of $\Delta v$ Approximation and Its Gradient

The sampling data are generated through the following procedure. First, the initial orbital elements of the spacecraft and the orbital differences between initial and target orbits are generated randomly within the ranges specified in Table 1. Note that $e_x$ and $e_y$ denote the two components of the eccentricity vector calculated by $e_x = e\cos\omega$ and $e_y = e\sin\omega$. The initial mass is randomly generated between 1,000 kg and 3,000 kg, while the transfer duration varies randomly from 60 to 300 days. The fixed thrust of 0.6 N maintains acceleration variability due to the randomly generated initial mass, with a specific impulse of $I_{sp}$ = 4,000 s. The generated data are filtered using the Lambert's solution: only datasets with impulsive velocity increments below 8,000 m/s are retained. Then, both the proposed approximation method and the indirect method [27] (as the benchmark for comparison) are employed to calculate equivalent velocity increments and their gradients.

**Table 1 Parameter ranges**

| Name of parameters | Range |
|---|---|
| Initial semimajor axis (AU) | [2.2, 2.8] |
| Initial $e_x$ | [0, 0.015] |
| Initial $e_y$ | [0, 0.015] |
| Initial inclination (deg) | [0, 5] |
| Initial right ascension of ascending node (deg) | [0, 360] |
| Initial argument of latitude (deg) | [0, 360] |
| Relative semimajor axis (AU) | [-0.5, 0.5] |
| Relative $e_x$ | [-0.015, 0.015] |
| Relative $e_y$ | [-0.015, 0.015] |
| Relative inclination (deg) | [-5, 5] |
| Relative right ascension of ascending node (RAAN, deg) | [0, 360] |
| Relative argument of latitude (deg) | [-10, 10] |
| Initial mass (kg) | [1,000, 3,000] |
| Transfer duration (day) | [50, 300] |

This process is repeated to generate 80,000 valid datasets. The iterative procedure terminates when the relative deviation between consecutive iterations falls below 0.1%, typically requiring 6 iterations on average. Fig. 1 presents the distribution of relative errors between approximate $\Delta v$ and benchmark results. It's demonstrated that the mean relative error (MRE) is 0.8% and the maximum error remains around 10%.

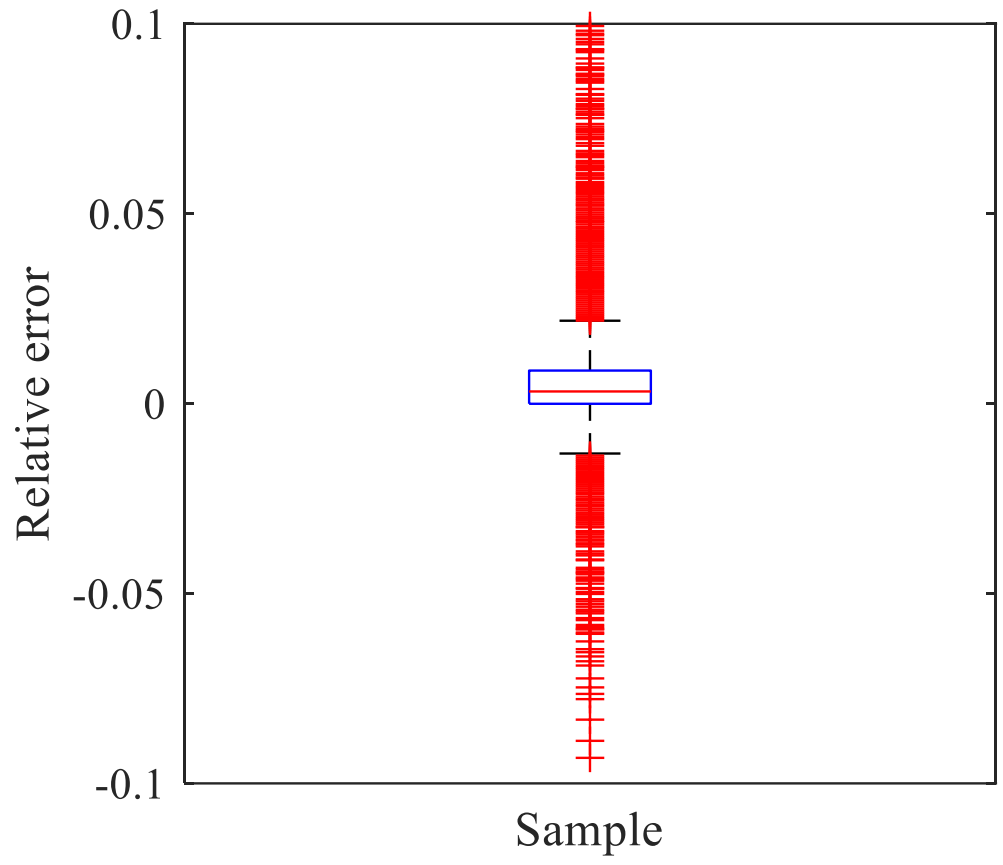

**Fig.1 Distribution of relative errors**

**Table 2 Orbital elements of a single case**

| | Semimajor axis (AU) | Eccentricity | Inclination (deg) | RAAN (deg) | Argument of perigee (deg) | Mean anomaly (deg) |
|---|---|---|---|---|---|---|
| Initial | 2.767 | 0.0402 | 1.71 | 110.17 | 178.58 | 199.0514 |
| Target | 2.764 | 0.0542 | 2.36 | 117.28 | 281.82 | 80.57043 |

For a single transfer in the data set with initial and target orbital elements listed in Table 2 and an initial spacecraft mass of 2,204 kg, the indirect method [27] yields a minimum transfer time of 149.8 days. Then, Fig. 2 compares the equivalent velocity increments obtained by both the proposed approximation method and the indirect method for transfer durations ranging from 149.8 to 300 days at 1-day intervals. It can be seen that the approximation results are slightly larger compared to the benchmark when $\Delta t$ is close to the minimum transfer time. However, the deviation becomes negligible for greater $\Delta t$. This behavior occurs because the approximation method equates impulsive maneuvers with fixed-direction continuous thrust arcs, which causes efficiency loss compared to the theoretically optimal thrust direction. Such efficiency loss would be more significant for near minimum-time transfers.

Notably, such error distribution (slightly more conservative feasibility assessment) actually facilitates more reliable conversion of approximate transfer arcs in the sequence into high-precision trajectories. It can well prevent situations where the approximation method might indicate the transfer is feasible, while the indirect method or other numerical low-thrust optimization algorithms fail to converge.

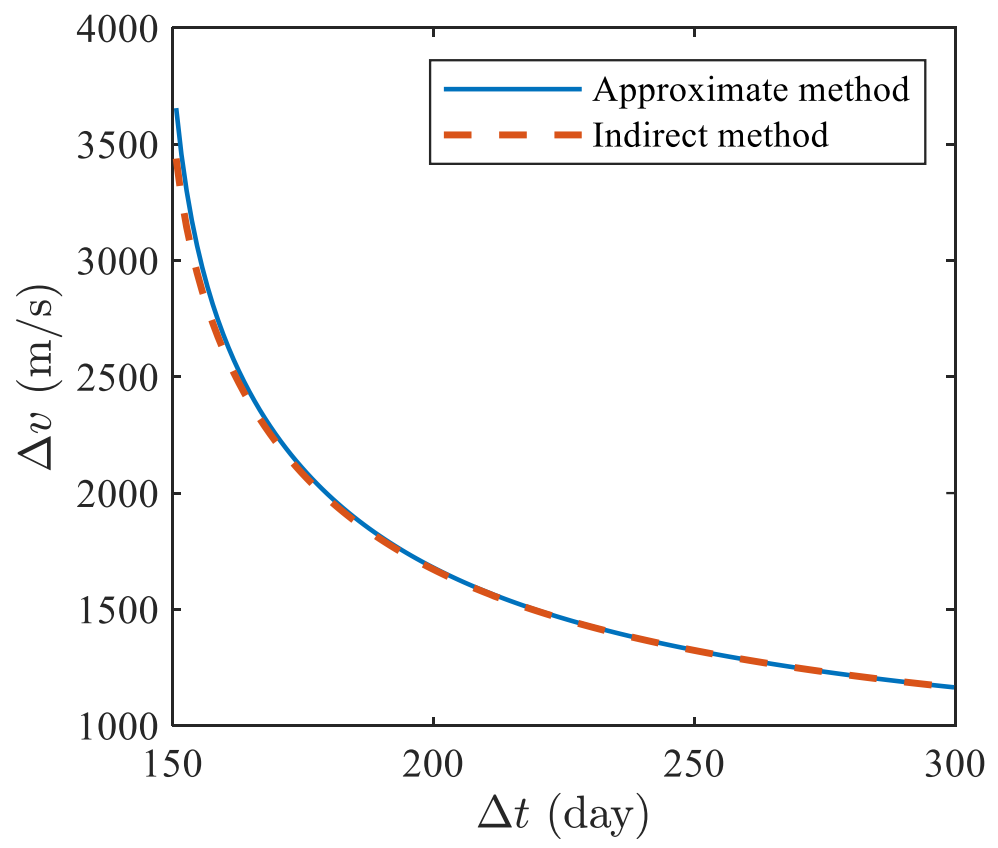

**Fig. 2 Comparison of $\Delta v$ between approximation and benchmark results**

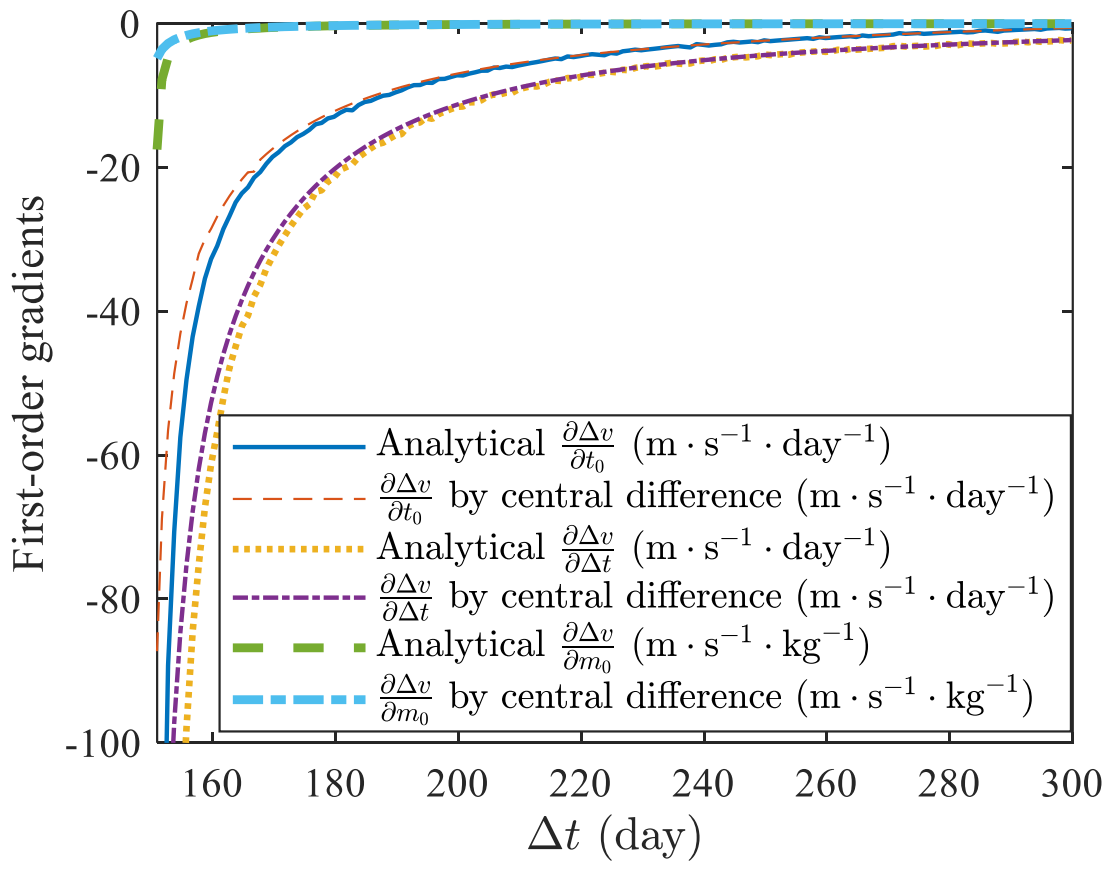

**Fig. 3 Comparison of first-order gradients between analytical and central difference results**

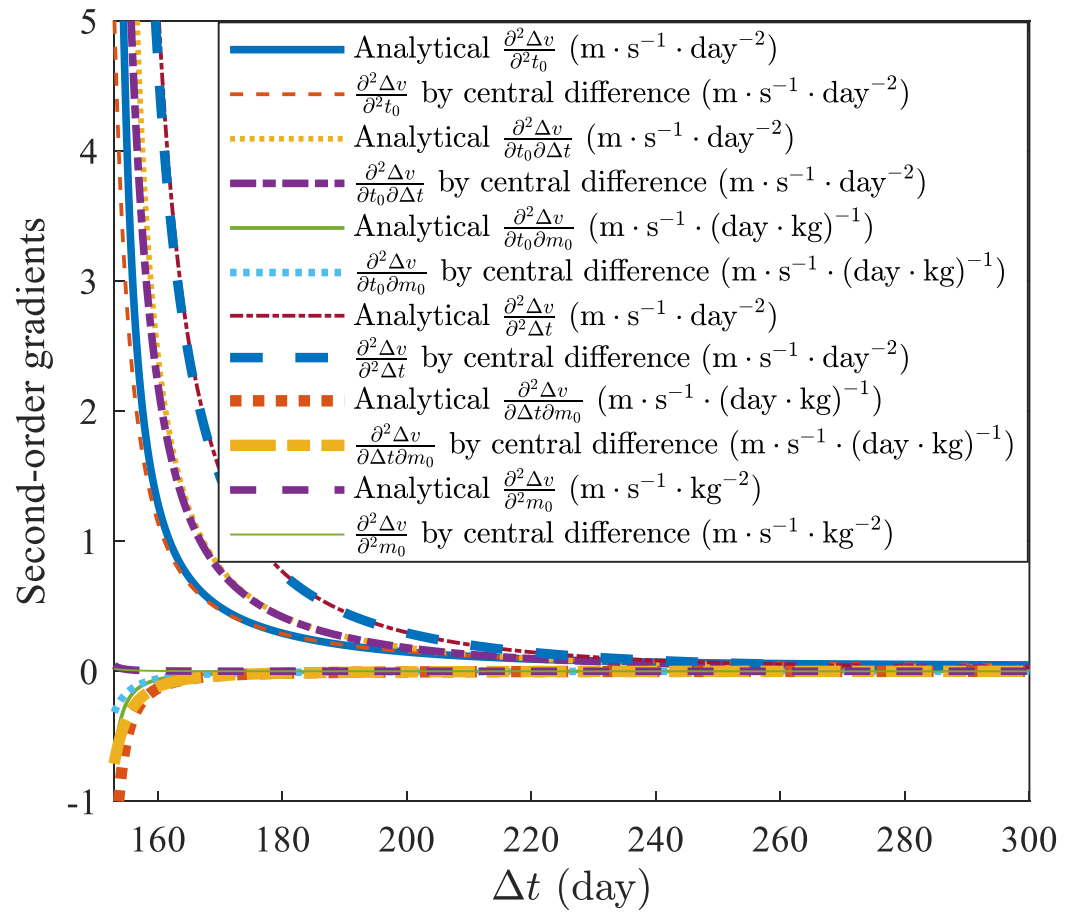

**Fig.4 Comparison of second-order gradients between approximation and central difference results**

Fig.3 compares the analytical first-order gradients, and Fig.4 compares the analytical second-order gradients from the approximation method with central-difference gradients from the indirect method, demonstrating the accuracy of the analytical gradients with different begin epochs and transfer durations. The proposed method requires an average computation time of $2.0\times10^{-5}$ s per solution on a desktop computer (CPU: 5.0 GHz), including $0.1\times10^{-5}$ s for the first-order gradient calculation and $0.4\times10^{-5}$ s for the second-order gradient calculation. In contrast, the indirect method [27] necessitates multiple random initial guesses and consumes over 0.05 s per solution. The MIMA [19] requires less computation time, but the accuracy is lower (showing 7.86 kg mean mass deviation versus 1.46 kg in our method). The approach of directly applying Lambert's method [15] with correction factors yields over 10% mean error and causes a significantly higher reachability misjudgment rate. This demonstrates that the proposed method achieves a balance between computational efficiency and accuracy for velocity increment estimation while providing direct access to analytical gradients.

**B. Nonlinear Programming of Multi-Asteroid Rendezvous Cases**

This subsection tests the proposed NLP algorithm using a 9-asteroid sequence. The orbital elements of targets can be queried in the data file provided by [17] using the target indices (9929-30125-18166-18219-25962-24838-56139-6291-24210). The initial mass is 2,500 kg, with a thrust of 0.6 N, specific impulse of 4,000 s, and a 40 kg toolkit deployed at each asteroid. The start time is fixed to $t_0$ = 622.0 days (where the times are given relative to Modified Julian Date of 64328). Using the branch and bound algorithm [5,11,17], a suboptimal feasible sequence with $\mathbf{x}_0$ = [0, 0.288, 0.222, 0.333, 0.333, 0.333, 0.222, 0.111, 0.6] has been obtained by setting $\Delta t_{\min}$ = 50 days and $\Delta t_{\max}$ = 450 days. $\mathbf{x}_0$ will be used as the initial value of SQP.

For the fuel-optimal problem, $t_f$ in the constraint function is set to 2,200 days. Then, convergence is achieved after 5 iterations when the norm of the increment in **x** falls below $1\times10^{-6}$, and $\mathbf{x}^*$ = [0.000, 0.355, 0.225, 0.399, 0.336, 0.351, 0.096, 0.204, 0.651]. The results are detailed in Table 3, which also includes the actual results obtained by fixing the arrival epochs and solving each transfer via the indirect method [27]. It's indicated that the estimated $\Delta v$ closely matches the actual $\Delta v$, with final mass deviations below 1.1 kg.

**Table 3 Fuel-optimal solution**

| Begin Target | Arrival Target | Begin time (day) | Arrival Time (day) | Approximate $\Delta v$ (m/s) | Approximate mass (kg) | Actual $\Delta v$ (m/s) | Actual mass (kg) |
|---|---|---|---|---|---|---|---|
| 9929 | 30125 | 622.00 | 831.59 | 1655.66 | 2356.68 | 1655.78 | 2356.67 |
| 30125 | 18166 | 831.59 | 982.75 | 1474.90 | 2229.71 | 1470.48 | 2229.96 |
| 18166 | 18219 | 982.75 | 1212.36 | 2068.71 | 2075.17 | 2062.53 | 2075.74 |
| 18219 | 25962 | 1212.36 | 1413.63 | 1839.84 | 1940.09 | 1839.02 | 1940.67 |
| 25962 | 24838 | 1413.63 | 1621.78 | 1893.02 | 1808.68 | 1889.88 | 1809.39 |
| 24838 | 56139 | 1621.78 | 1715.18 | 1665.05 | 1693.52 | 1653.05 | 1694.72 |
| 56139 | 6291 | 1715.18 | 1857.15 | 990.57 | 1611.29 | 990.40 | 1612.47 |
| 6291 | 24210 | 1857.15 | 2200.00 | 1735.18 | 1501.57 | 1736.17 | 1502.66 |

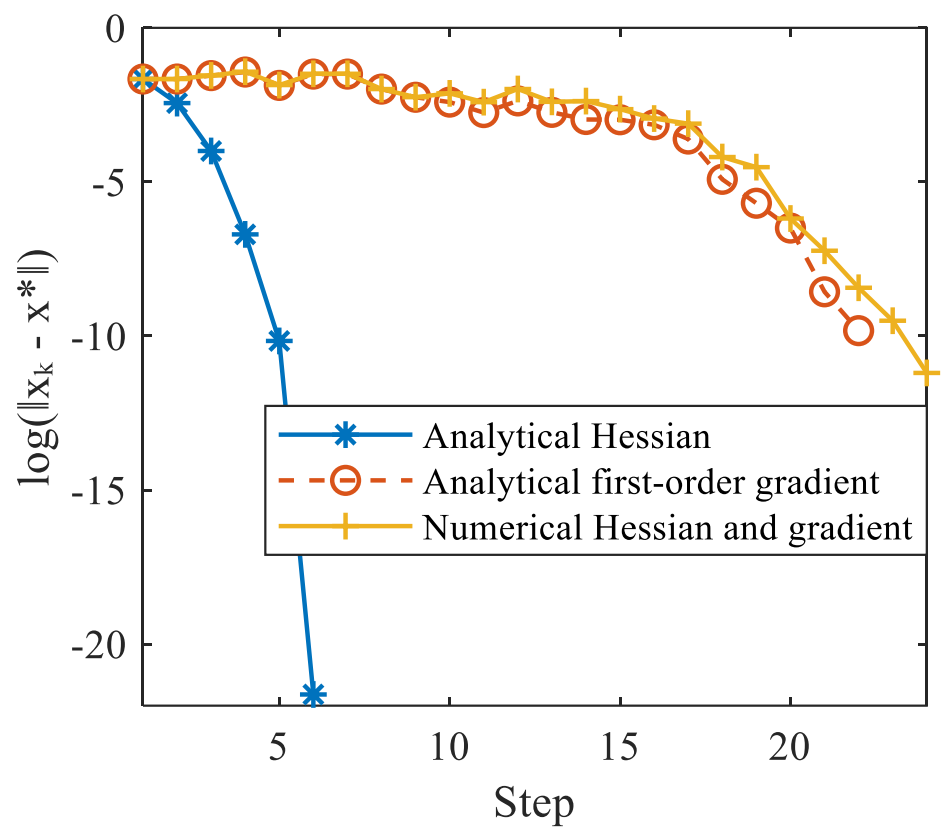

**Fig.5 Convergence comparison for fuel-optimal problem**

The proposed algorithm required fewer than 35 times of objective function evaluations (about 0.02 seconds), and Fig.5 illustrates the curves of $\log(\|x_k - x^*\|)$ against iteration steps. Using the convergence rate defined by

$$p = \frac{\log(\|\mathbf{x}_{k+1} - \mathbf{x}^*\| / \|\mathbf{x}_k - \mathbf{x}^*\|)}{\log(\|\mathbf{x}_k - \mathbf{x}^*\| / \|\mathbf{x}_{k-1} - \mathbf{x}^*\|)}, \tag{34}$$

the average convergence rate of the last five steps is 2.0, which is consistent with the theoretical performance of SQP. Fig. 5 also presents the convergence histories obtained by 'fmincon' function in MATLAB for two other cases: using analytical first-order gradients and using numerical gradients, which require 122 and 333 times of objective function evaluations, respectively (the average convergence rates are 1.5 and 1.6, respectively). This demonstrates the effectiveness of the analytical Hessian derived in this Note. Additionally, the differential evolution algorithm [7, 28] is also tested, which requires over 20,000 objective function calls, yet the best solution yields negligible differences in terminal mass and arriving epochs compared to our results.

**Table 4 Time-optimal solution**

| Begin Target | Arrival Target | Begin time (day) | Arrival Time (day) | Approximate $\Delta v$ (m/s) | Approximate mass (kg) | Actual $\Delta v$ (m/s) | Actual mass (kg) |
|---|---|---|---|---|---|---|---|
| 9929 | 30125 | 622.00 | 816.71 | 1843.38 | 2345.24 | 1843.65 | 2345.22 |
| 30125 | 18166 | 816.71 | 958.86 | 1637.86 | 2209.33 | 1630.00 | 2209.76 |
| 18166 | 18219 | 958.86 | 1165.60 | 2441.64 | 2036.00 | 2430.99 | 2036.97 |
| 18219 | 25962 | 1165.60 | 1350.12 | 2363.84 | 1876.93 | 2363.66 | 1877.86 |
| 25962 | 24838 | 1350.12 | 1501.81 | 2197.60 | 1734.67 | 2193.12 | 1735.75 |
| 24838 | 56139 | 1501.81 | 1653.31 | 2783.00 | 1575.87 | 2753.26 | 1578.10 |
| 56139 | 6291 | 1653.31 | 1734.93 | 1959.87 | 1459.07 | 1942.92 | 1461.84 |
| 6291 | 24210 | 1734.93 | 1861.17 | 3350.31 | 1299.62 | 3320.87 | 1303.17 |

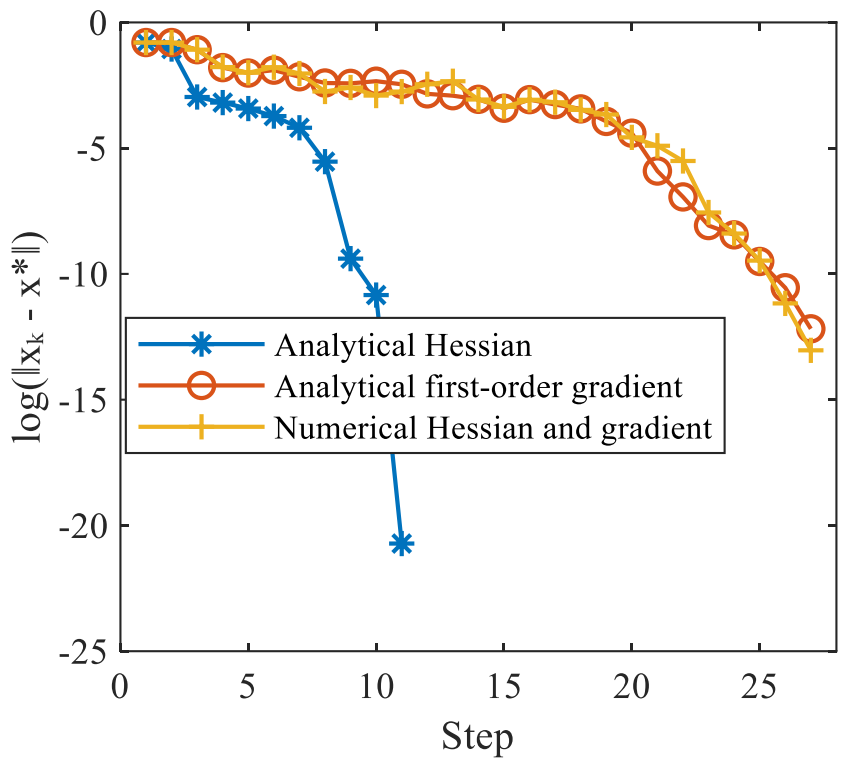

**Fig.6 Convergence comparison for time-optimal problem**

For the time-optimal problem, we set the minimal terminal mass $m_f$ = 1,300 kg. Then, the obtained optimal $t_f$ is 1,861.17 days, and the results are detailed in Table 4. The proposed algorithm required fewer than 60 times of objective function evaluations. By contrast, 'fmincon' required 142 and 366 times of objective function evaluations when using the analytical first-order gradients and using numerical gradients, respectively. Fig.6 illustrates the curves of log(‖$x_k$ - x*‖) against iteration steps, which indicates that the proposed algorithm performs equally well in time-optimal scenarios.

In summary, the NLP algorithm proposed in this Note is more suitable as a local refinement algorithm for pre-constructed feasible solutions (e.g., those obtained through fixed-step branch-and-bound methods). The computational cost is acceptable for integration into global optimization algorithms to improve sequences. It should be noted that in cases where feasible initial values are unavailable, optimal solutions can be obtained through repeated random guesses and iterative solving. We also test random initial guesses, requiring about 1,000 trials on average to find a feasible initial solution. In this condition, the computational cost becomes heavier, thereby diminishing the efficiency advantage over existing evolutionary algorithms.

### C. Nonlinear Programming of GTOC12 Cases

In the GTOC12 problem [17], multiple spacecraft are required to jointly perform two rendezvous with selected targets (from 60,000 candidate asteroids), to maximize the total mining mass proportional to the sum of time intervals between two rendezvous of each selected target (36.525 kg per year). Each spacecraft is required to depart from Earth to start and finally return to Earth to complete the mission. In this Note, only the transfers between asteroids are tested to make comparison because the proposed $\Delta v$ approximator is inapplicable for Earth-asteroid transfers.

Then, each rendezvous sequence can be divided into two subsequences: the first rendezvous phase to deploy the mining toolkit (40 kg) and the second rendezvous phase to collect the mined mass. Though the selected targets are allowed to be cooperatively visited twice using different spacecrafts, however, for any individual spacecraft, advancing the rendezvous epochs in subsequence 1 (first rendezvous) and delaying the rendezvous epochs in subsequence 2 (second rendezvous) both positively contribute to the overall mined mass. Therefore, maximizing the sum of rendezvous epochs of subsequence 2 and minimizing the sum of rendezvous epochs of subsequence 1 for a single-spacecraft sequence with fixed target is consistent with the original objective function given by GTOC12. The objective function for the single-spacecraft optimization in this section can thus be expressed by

$$J = \sum_{k=1}^{n_2} t_{k,2} - \sum_{k=1}^{n_1} t_{k,1} \tag{35}$$

where $n_1$ and $n_2$ are the target number of two subsequences, $t_{k,1}$ represent the rendezvous epochs in subsequence 1 and $t_{k,2}$ represent the rendezvous epochs in subsequence 2. The gradient of Eq. (35) is:

$$\sum_{k=1}^{n_2} \frac{\partial t_{k,2}}{\partial x_i} - \sum_{k=1}^{n_1} \frac{\partial t_{k,1}}{\partial x_i} \tag{36}$$

In this scene, both time and fuel constraints (Eqs. (4) and (6)) should be considered (denoted by $g_1$ and $g_2$). The feasibility check (Eq. (23)) should also be included (denoted by $g_i$, $i$ = 2,…,n). Therefore, the number of constraints is $n$+1. The gradients of constraint functions have been derived in Section IV, and the Hessian matrix is modified to:

$$\nabla^2 L(x_k, \mu) = -\frac{\mu_0 \partial^2 m_n}{\partial x_k \partial x_j} + \sum_{i=2}^{n} \frac{\mu_i \partial^2 g_i}{\partial x_k \partial x_j} \tag{37}$$

The validation was conducted using solutions submitted by the top three teams in GTOC12 (the first spacecraft of JPL submission [4], the first spacecraft of BIT submission [18], and the second spacecraft of CS submission [23]). As the transfers leaving and returning to Earth are excluded in this section, $t_0$ and $t_f$ are set equal to the arrival time of the first asteroid and the departure time of the last asteroid in the original submissions, respectively. $m_0$ and $m_f$ are set in the same manner. Their values are detailed in Table 5.

It should be noted that the mined mass from each asteroid may change with the corresponding rendezvous epochs during the optimization. Here, it's assumed that mass changes compared to those in the original submissions are small, and according to Eq. (10), the changes in the average accelerations are negligible. Then, the collected masses ($m_{kit}$ in subsequence 2) added to the spacecraft after each rendezvous are fixed to be the same as those in the original submissions. Thus, $m_{kit}$ is set to 40 kg for subsequence 1, and for subsequence 2, $m_{kit}$ is set to the negative values of mined masses of target asteroid in the original submissions. The solving process remains the same as that proposed in Sections IV and V.B.

**Table 5 Details of the original three sequences**

| Sequence | Target indices | $t_0$ (day) | $t_f$ (day) | $m_0$ (kg) | $m_f$ (kg) |
|---|---|---|---|---|---|
| The first sequence of JPL submission | Sub sequence 1: 12095, 3506, 49192, 33590, 36666, 2154, 33908, 35666, 4971<br>Sub sequence 2: 4971, 11519, 41936, 6728, 27741, 37789, 51492, 51006, 12095 | 546.28 | 4924.55 | 2496.62 | 1417.11 |
| The first sequence of BIT submission | Sub sequence 1: 29204, 11732, 58731, 45541, 24461<br>Sub sequence 2: 24461, 11380, 9769, 59982, 57619, 33590, 36739, 49192, 36666, 28411, 8913, 21776 | 584 | 4964 | 2546.84 | 1867.86 |
| The second sequence of CS submission | Sub sequence 1: 510, 10040, 26270, 9914, 11267, 14589, 7579, 49312<br>Sub sequence 2: 49312, 14589, 10040, 510, 26270, 11267, 7579, 9914 | 540.02 | 4915.67 | 2480.65 | 1395.32 |

The re-optimized arrival epochs of three sequences are presented in Tables 6, 7, and 8, respectively, where each transfer leg has been solved via the indirect method to obtain the actual masses. It's seen that the mass before returning to Earth of each sequence is greater than that in the original submissions (the fuel constraint is satisfied), and the rendezvous epoch of the last asteroid of each sequence is also not later than that in the original submissions (the time constraint is satisfied). The advanced rendezvous epochs in subsequence 1 and delayed rendezvous epochs in subsequence 2 improve the total mined mass of three sequences by 4.64 kg, 13.91 kg, and 2.76 kg, respectively, demonstrating that the NLP algorithm achieves improved metrics under the same fuel and time constraints. It's proven that the method exhibits better local optimization capability.

**Table 6 Re-optimized result of JPL submission**

| Begin Target | Arrival Target | Begin time (day) | Arrival Time (day) | Actual mass (kg) |
|---|---|---|---|---|
| 12095 | 3506 | 546.28 | 850.20 | 2190.30 |
| 3506 | 49192 | 850.20 | 1051.24 | 2053.74 |
| 49192 | 33590 | 1051.24 | 1106.01 | 1958.57 |
| 33590 | 36666 | 1106.01 | 1306.00 | 1792.64 |
| 36666 | 2154 | 1306.00 | 1466.90 | 1664.13 |
| 2154 | 33908 | 1466.90 | 1798.76 | 1507.70 |
| 33908 | 35666 | 1798.76 | 2113.49 | 1404.66 |
| 35666 | 4971 | 2113.49 | 2354.71 | 1301.79 |
| 4971 | 4971 | 2354.71 | 3212.64 | 1323.01 |
| 4971 | 11519 | 3212.64 | 3714.44 | 1282.26 |
| 11519 | 41936 | 3714.44 | 3972.13 | 1320.00 |
| 41936 | 6728 | 3972.13 | 4082.42 | 1360.12 |
| 6728 | 27741 | 4082.42 | 4280.30 | 1378.15 |
| 27741 | 37789 | 4280.30 | 4578.11 | 1333.38 |
| 37789 | 51492 | 4578.11 | 4690.45 | 1378.28 |
| 51492 | 51006 | 4690.45 | 4788.72 | 1403.05 |
| 51006 | 12095 | 4788.72 | 4924.54 | 1420.37 |

**Table 7 Re-optimized result of BIT submission**

| Begin Target | Arrival Target | Begin time (day) | Arrival Time (day) | Actual mass (kg) |
|---|---|---|---|---|
| 29204 | 11732 | 585.901 | 844.239 | 2408.05 |
| 11732 | 58731 | 844.239 | 899.946 | 2294.77 |
| 58731 | 45541 | 899.946 | 1064.59 | 2132.98 |
| 45541 | 24461 | 1064.59 | 1290.8 | 2034 |
| 24461 | 24461 | 1290.8 | 3257.39 | 2075.09 |
| 24461 | 11380 | 3257.39 | 3536.23 | 2021.3 |
| 11380 | 9769 | 3536.23 | 3704.47 | 1976.82 |
| 9769 | 59982 | 3704.47 | 3846.98 | 1939.48 |
| 59982 | 57619 | 3846.98 | 3978.24 | 1942.69 |
| 57619 | 33590 | 3978.24 | 4123.84 | 1913.87 |
| 33590 | 36739 | 4123.84 | 4269.9 | 1892.16 |
| 36739 | 49192 | 4269.9 | 4434.33 | 1874.42 |
| 49192 | 36666 | 4434.33 | 4587.17 | 1857.8 |
| 36666 | 28411 | 4587.17 | 4637.54 | 1893.43 |
| 28411 | 8913 | 4637.54 | 4799.34 | 1879.13 |
| 8913 | 21776 | 4799.34 | 4963.97 | 1868.83 |

**Table 8 Re-optimized result of CS submission**

| Begin Target | Arrival Target | Begin time (day) | Arrival Time (day) | Actual mass (kg) |
|---|---|---|---|---|
| 510 | 10040 | 540.053 | 771.635 | 2267.59 |
| 10040 | 26270 | 771.635 | 907.038 | 2101.4 |
| 26270 | 9914 | 907.038 | 1088.71 | 1918.86 |
| 9914 | 11267 | 1088.71 | 1298.98 | 1756.24 |
| 11267 | 14589 | 1298.98 | 1509.63 | 1601.57 |
| 14589 | 7579 | 1509.63 | 1926.36 | 1435.06 |
| 7579 | 49312 | 1926.36 | 2245.85 | 1325.95 |
| 49312 | 49312 | 2245.85 | 3925.76 | 1371.46 |
| 49312 | 14589 | 3925.76 | 4044.14 | 1399.34 |
| 14589 | 10040 | 4044.14 | 4143.85 | 1437.32 |
| 10040 | 510 | 4143.85 | 4319.6 | 1442.21 |
| 510 | 26270 | 4319.6 | 4460.58 | 1445.16 |
| 26270 | 11267 | 4460.58 | 4718.08 | 1401.93 |
| 11267 | 7579 | 4718.08 | 4829.23 | 1374.61 |
| 7579 | 9914 | 4829.23 | 4915.66 | 1396.15 |

## VI. Conclusion

This Note presents an analytical formulation for both first- and second-order gradients of low-thrust rendezvous $\Delta v$ with respect to the start epoch, transfer duration, and initial mass, derived through an iterative Lambert-based $\Delta v$ estimator. These gradients are subsequently utilized to construct the analytical Hessian and formulate a sequential quadratic programming (SQP) framework for optimizing multi-rendezvous low-thrust trajectories. Simulation results demonstrate that the mean relative error of $\Delta v$ approximation remains below 0.8% for transfers between asteroids in the main belt, with the analytical gradients aligning closely with those computed via the central difference method. The effectiveness of the analytical Hessian is further validated through a 9-asteroid rendezvous sequence under both fuel-optimal and time-optimal objectives, where the SQP algorithm exhibits significantly improved convergence performance. Additionally, the method is applied to three sequences from the top-ranking submissions in the 12$^{th}$ Global Trajectory Optimization Competition (GTOC12), consistently improving their original metrics. The proposed approach is well-suited for integration into global trajectory optimization algorithms for multi-spacecraft multi-target missions, offering enhanced accuracy in objective function evaluation while maintaining high computational efficiency.

## Acknowledgement

The work was supported by the National Natural Science Foundation of China (No. 12202504).